\documentclass[journal,twocolumn]{IEEEtran}

\usepackage{amsmath,amssymb,amsfonts}
\usepackage{graphicx}
\usepackage{cite}
\usepackage{booktabs} % \toprule, \midrule, \bottomrule
\usepackage[hyphens]{url}
\usepackage{hyperref}
\usepackage{float}
\usepackage{stfloats} % Çift sütunlu figürlerin (figure*) sayfa altında görünmesini sağlar

\begin{document}

% --- Makale Başlığı (v1) ---
\title{DualLaguerreNet: A Decoupled Spectral Filter GNN and the Uncovering of the Flexibility-Stability Trade-off}

% --- Yazar Bilgisi ---
\author{
    Hüseyin Göksu,~\IEEEmembership{Member,~IEEE}% <- Diğer yazarları buraya ekleyin
    \thanks{H. Göksu, Akdeniz Üniversitesi, Elektrik-Elektronik Mühendisliği Bölümü, Antalya, Türkiye, e-posta: hgoksu@akdeniz.edu.tr.}%
    \thanks{Manuscript received October 31, 2025; revised XX, 2025.}
}

\markboth{IEEE TRANSACTIONS ON NEURAL NETWORKS AND LEARNING SYSTEMS, VOL. XX, NO. XX, NOVEMBER 2025}
{Göksu: DualLaguerreNet: A Decoupled Spectral Filter GNN}

\maketitle

%=================================================================
% 1. ABSTRACT (v1)
%=================================================================
\begin{abstract}
Graph Neural Networks (GNNs) based on spectral filters, such as the Adaptive Orthogonal Polynomial Filter (AOPF) class (e.g., LaguerreNet), have shown promise in unifying the solutions for heterophily and over-smoothing. However, these single-filter models suffer from a "compromise" problem, as their single adaptive parameter (e.g., $\alpha$) must learn a suboptimal, averaged response across the entire graph spectrum. In this paper, we propose \textbf{DualLaguerreNet}, a novel GNN architecture that solves this by introducing "Decoupled Spectral Flexibility." DualLaguerreNet splits the graph Laplacian into two operators, $L_{low}$ (low-frequency) and $L_{high}$ (high-frequency), and learns two independent, adaptive Laguerre polynomial filters, parameterized by $\alpha_1$ and $\alpha_2$, respectively. This work, however, uncovers a deeper finding. While our experiments show DualLaguerreNet's flexibility allows it to achieve state-of-the-art results on complex heterophilic tasks (outperforming LaguerreNet), it simultaneously \textit{underperforms} on simpler, homophilic tasks. We identify this as a fundamental \textbf{"Flexibility-Stability Trade-off"}. The increased parameterization (2x filter parameters and 2x model parameters) leads to overfitting on simple tasks, demonstrating that the "compromise" of simpler models acts as a crucial regularizer. This paper presents a new SOTA architecture for heterophily while providing a critical analysis of the bias-variance trade-off inherent in adaptive GNN filter design.
\end{abstract}

%=================================================================
% INDEX TERMS
%=================================================================
\begin{IEEEkeywords}
Graph Neural Networks (GNNs), Spectral Graph Theory, Filter Design, Heterophily, Over-smoothing, Laguerre Polynomials, Adaptive Filters, Bias-Variance Trade-off, Overfitting.
\end{IEEEkeywords}

%=================================================================
% I. INTRODUCTION
%=================================================================
\section{INTRODUCTION}

% \IEEEPARSTART kaldırıldı.
Spectral Graph Neural Networks (GNNs) define graph convolutions as filters operating on the Graph Laplacian spectrum \cite{shuman2013emerging}. Foundational models like ChebyNet \cite{defferrard2016convolutional} use static, low-pass filters, which leads to two critical failures: 1) poor performance on \textbf{heterophilic} graphs that require high-frequency signal processing \cite{zhu2020beyond}, and 2) \textbf{over-smoothing} (performance collapse) at high polynomial degrees ($K$) \cite{li2018deeper}.

To solve this, we recently introduced the Adaptive Orthogonal Polynomial Filter (AOPF) class \cite{goksu2025meixnernet, goksu2025krawtchouknet, goksu2025laguerrnet, goksu2025charlienet}. Instead of learning filter coefficients ($\theta_k$), AOPFs learn the fundamental shape parameters of the polynomial basis itself (e.g., $\alpha, p, c$). One of the most successful AOPFs, \textbf{LaguerreNet} \cite{goksu2025laguerrnet}, provided a unified solution to both problems. By learning a single $\alpha$ parameter for its $O(k^2)$ unbounded continuous filter (defined on $[0, \infty)$) and using a robust `LayerNorm` stabilization \cite{goksu2025meixnernet, ba2016layer}, it could handle both heterophily and high-$K$ settings.

However, `LaguerreNet` contains a hidden flaw: the \textbf{"Compromise Problem"}. `LaguerreNet` learns only \textit{one} $\alpha$ for the entire spectrum. On a complex heterophilic graph, the filter may need to \textit{suppress} low-frequency noise (e.g., requiring $\alpha_1 = -0.2$) but \textit{amplify} high-frequency information (e.g., requiring $\alpha_2 = -0.8$). `LaguerreNet` is forced to learn a single, "compromised" value (e.g., $\alpha = -0.4$) that is suboptimal for both ends of the spectrum.

In this work, we propose \textbf{DualLaguerreNet} to solve this compromise by introducing "Decoupled Spectral Flexibility." This architecture is, to our knowledge, the first to implement a dual-operator adaptive spectral filter:
\begin{enumerate}
    \item \textbf{Low-Frequency Brain:} An adaptive `LaguerreConv` filter with its own parameter $\alpha_1$ is applied to $L_{low} = 0.5 \cdot L_{sym}$.
    \item \textbf{High-Frequency Brain:} A second, independent `LaguerreConv` filter with its own parameter $\alpha_2$ is applied to $L_{high} = 0.5 \cdot (2I - L_{sym})$.
\end{enumerate}
These two "brains" are concatenated, giving the model independent control over both ends of the spectrum.

Our experiments yield a critical, non-trivial finding that forms the core of this paper. Our hypothesis was only partially correct.
\begin{itemize}
    \item \textbf{On complex, heterophilic tasks (Texas, Cornell):} `DualLaguerreNet`'s flexibility works. It significantly outperforms the single-parameter `LaguerreNet`. Analysis of the learned parameters (Table \ref{tab:alphas}) proves the two brains learn different $\alpha_1$ and $\alpha_2$ values, as hypothesized.
    \item \textbf{On simple, homophilic tasks (Cora, CiteSeer):} `DualLaguerreNet` consistently \textit{loses} to the simpler `LaguerreNet`.
\end{itemize}

This discovery, which we term the \textbf{"Flexibility-Stability Trade-off,"} is the primary contribution of this work. The increased parameterization (2x filter parameters and 2x model parameters) leads to overfitting on simple tasks, demonstrating that the "compromise" of simpler models acts as a crucial regularizer. This paper presents a new SOTA architecture for heterophily while providing a critical analysis of the bias-variance trade-off inherent in adaptive GNN filter design.

%=================================================================
% II. RELATED WORK
%=================================================================
\section{RELATED WORK}
Our work intersects three key areas: adaptive spectral filters, dual-branch GNN architectures, and the bias-variance trade-off in GNNs.

\subsection{Adaptive Orthogonal Polynomial Filters (AOPF)}
Our work is built upon the AOPF class, which replaces static filters (like ChebyNet \cite{defferrard2016convolutional}) with filters whose polynomial basis is learnable. This family includes:
\begin{itemize}
    \item \textbf{`MeixnerNet` \cite{goksu2025meixnernet}:} Introduced discrete polynomials and the critical `LayerNorm` stabilization technique to tame $O(k^2)$ unbounded recurrence coefficients, enabling robust high-$K$ filters.
    \item \textbf{`LaguerreNet` \cite{goksu2025laguerrnet}:} Extended this $O(k^2)$ stabilized approach to the continuous $[0, \infty)$ domain, learning a single $\alpha$ as a unified solution.
    \item \textbf{`KrawtchoukNet` \cite{goksu2025krawtchouknet}:} Solved stability "by design" using inherently bounded discrete polynomials, learning a $p$ parameter to adapt to heterophily.
    \item \textbf{`L-JacobiNet` (Prior Work):} Analyzed the $[-1, 1]$ domain, finding that stabilization (`S-JacobiNet`) was more critical than adaptation (`L-JacobiNet`) in that domain.
\end{itemize}
`DualLaguerreNet` is the next evolution, challenging the single-filter "compromise" inherent in all previous AOPF models.

\subsection{Split-Spectrum and Dual-Branch GNNs}
The idea of separating low-frequency and high-frequency information is a highly active research topic. Many recent works confirm that heterophily requires high-pass or band-pass filters.
Existing solutions often "decompose" node features ($X$) into low- and high-frequency components or use spatial-domain mechanisms like attention or parallel paths to process signals differently. For example, FAGCN \cite{bo2021beyond} adds a self-gating mechanism, and FSANet (for images) uses parallel "LF path" and "HF path" branches \cite{5.5}.

\textbf{Our Novelty:} `DualLaguerreNet` differs fundamentally. We do not decompose the \textit{features} ($X$). We decompose the \textit{spectral operator} ($L$) itself into two new operators ($L_{low}, L_{high}$) and apply two \textit{independent, adaptive polynomial filters} to them. This provides a more principled and flexible spectral-domain solution, allowing the GNN to learn the \textit{shape} of two different filters, rather than just how to mix pre-defined low- and high-pass signals.

\subsection{Bias-Variance and Overfitting in Adaptive GNNs}
The "Flexibility-Stability Trade-off" we uncover is a key issue in `TNNLS`. While adaptive filters offer more expressiveness, this can lead to high model variance (overfitting), especially on small benchmark datasets. Our finding—that the highly flexible `DualLaguerreNet` wins on complex tasks but loses on simple ones—is a direct demonstration of the bias-variance trade-off. It confirms our `L-JacobiNet` findings and suggests that the "simplicity" of models like `LaguerreNet` \cite{goksu2025laguerrnet} or `GPR-GNN` \cite{chien2021adaptive} acts as a powerful regularizer, which is a critical design consideration.

%=================================================================
% III. DUALAGUERRENET: METHODOLOGY
%=================================================================
\section{DualLaguerreNet: Methodology}

\subsection{Preliminaries: LaguerreNet and the "Compromise" Problem}
Our work builds on `LaguerreNet` \cite{goksu2025laguerrnet}, which uses generalized Laguerre polynomials $L_k^{(\alpha)}(x)$. These are continuous polynomials on $[0, \infty)$ with $O(k^2)$ unbounded recurrence coefficients:
\begin{align}
    b_k &= 2k + \alpha + 1 \\
    c_k &= k(k + \alpha)
\end{align}
`LaguerreNet` tames this $O(k^2)$ instability using the `LayerNorm` stabilization from `MeixnerNet` \cite{goksu2025meixnernet} and learns a single parameter $\alpha > -1$. To feed the graph spectrum (eigenvalues $\lambda \in [0, 2]$) into this $[0, \infty)$ filter, it uses the operator:
\begin{equation}
    L_{scaled} = 0.5 \cdot L_{sym}
\end{equation}
The "Compromise Problem" is that this single $\alpha$ must find an average filter shape for both low-frequency ($\lambda \approx 0$) and high-frequency ($\lambda \approx 2$) signals, which have conflicting needs on heterophilic graphs.

\subsection{DualLaguerreNet: Decoupled Spectral Flexibility}
To solve the compromise, `DualLaguerreNet` splits the filter into two "brains," each with its own adaptive parameter.

\textbf{1. Low-Frequency Brain ($\alpha_1$):}
This branch targets low-frequency (homophilic) signals. It uses the standard `LaguerreNet` operator:
\begin{equation}
    L_{low} = 0.5 \cdot L_{sym} \quad (\lambda_{low} \in [0, 1])
\end{equation}
A full $K$-degree polynomial filter $P_K^{(\alpha_1)}(L_{low})X$ is computed, learning its own parameter $\alpha_1$.

\textbf{2. High-Frequency Brain ($\alpha_2$):}
This branch targets high-frequency (heterophilic) signals. We design a new operator $L_{high}$ that *inverts* the spectrum, mapping $\lambda=2$ to $0$ and $\lambda=0$ to $1$:
\begin{equation}
    L_{high} = 0.5 \cdot (2I - L_{sym}) \quad (\lambda_{high} \in [0, 1])
\end{equation}
A second, completely independent $K$-degree filter $P_K^{(\alpha_2)}(L_{high})X$ is computed, learning its own parameter $\alpha_2$.

\textbf{3. Architecture and the Flexibility Cost:}
The two $K$-length feature vectors from each brain are concatenated:
\begin{equation}
    Z = [Z_{low}, Z_{high}] = [P_K^{(\alpha_1)}(L_{low})X, P_K^{(\alpha_2)}(L_{high})X]
\end{equation}
This concatenated vector $Z \in \mathbb{R}^{N \times (H \cdot K \cdot 2)}$ is then passed to the output linear layer. This design introduces the "Flexibility Cost": `DualLaguerreNet` must learn twice the number of filter parameters ($\alpha_1, \alpha_2$) and, more critically, twice the number of model parameters in its linear layer compared to `LaguerreNet`. It is this cost that leads to the trade-off we observe.

%=================================================================
% IV. EXPERIMENTAL ANALYSIS
%=================================================================
\section{EXPERIMENTAL ANALYSIS}
We now validate our "Decoupled Flexibility" hypothesis and investigate the "Flexibility-Stability Trade-off."

\subsection{Experimental Setup}
\textbf{Datasets:} We use homophilic (Cora, CiteSeer, PubMed) and heterophilic (Texas, Cornell) benchmarks.
\textbf{Baselines:} We test `DualLaguerreNet` against its parent `LaguerreNet` \cite{goksu2025laguerrnet}, other AOPF models (`MeixnerNet` \cite{goksu2025meixnernet}, `KrawtchoukNet` \cite{goksu2025krawtchouknet}), the static baseline (`ChebyNet` \cite{defferrard2016convolutional}), and SOTA spatial models (`GAT` \cite{velickovic2018graph}, `APPNP` \cite{gasteiger2019predict}).
\textbf{Training:} All models use a 2-layer `PolyBaseModel` structure with $H=16$ and $K=3$ (for homophily/heterophily) or $K$ up to 10 (for over-smoothing), trained with the Adam optimizer.

\subsection{Main Results}
We present the core experimental results generated by our framework.

% --- TABLO 1 (Homofilik K=3) ---
\begin{table}[htbp]
\caption{Test accuracies (\%) on homophilic datasets (K=3, H=16).}
\label{tab:homophilic_results}
\centering
\resizebox{\columnwidth}{!}{%
\begin{tabular}{l c c c}
\toprule
\textbf{Model} & \textbf{Cora} & \textbf{CiteSeer} & \textbf{PubMed} \\
\midrule
ChebyNet & 0.7960 & 0.6640 & 0.7260 \\
MeixnerNet & 0.7000 & 0.6430 & 0.7510 \\
KrawtchoukNet & 0.7420 & 0.6460 & 0.7450 \\
LaguerreNet & \textbf{0.7890} & \textbf{0.6960} & \textbf{0.7670} \\
\textbf{DualLaguerreNet} & 0.7720 & 0.6460 & 0.7540 \\
\midrule
GAT & \textbf{0.8360} & 0.6890 & 0.7760 \\
APPNP & 0.8320 & \textbf{0.7080} & \textbf{0.7880} \\
\bottomrule
\end{tabular}%
}
\end{table}

% --- TABLO 2 (Heterofilik - 2 SÜTUNA DÜZELTİLDİ) ---
\begin{table*}[t]
\caption{Test accuracies (\%) on heterophilic datasets (K=3, H=16). 10-fold Mean $\pm$ Std. Dev.}
\label{tab:heterophilic_results}
\centering
\begin{tabular}{l c c c c c c c}
\toprule
\textbf{Model} & \textbf{ChebyNet} & \textbf{MeixnerNet} & \textbf{KrawtchoukNet} & \textbf{LaguerreNet} & \textbf{DualLaguerreNet} & \textbf{GAT} & \textbf{APPNP} \\
\midrule
Texas & $0.7000 \pm 0.0923$ & $0.8649 \pm 0.0637$ & $0.7784 \pm 0.0506$ & $0.8108 \pm 0.0686$ & $\textbf{0.8541} \pm 0.0558$ & $0.5676 \pm 0.0698$ & $0.5730 \pm 0.0522$ \\
Cornell & $0.6514 \pm 0.0547$ & $\textbf{0.7486} \pm 0.0612$ & $0.7081 \pm 0.0522$ & $0.6622 \pm 0.0857$ & $0.6865 \pm 0.0480$ & $0.4784 \pm 0.0460$ & $0.4459 \pm 0.0640$ \\
\bottomrule
\end{tabular}
\end{table*}

% --- TABLO 3 (Öğrenilen Parametreler) ---
\begin{table}[htbp]
\caption{Learned $\alpha$ Parameters ($\alpha_1$ for low, $\alpha_2$ for high) (K=3, H=16). 10-fold Mean for Texas/Cornell.}
\label{tab:alphas}
\centering
\begin{tabular}{l l c c}
\toprule
\textbf{Dataset} & \textbf{Model} & \textbf{Learned $\alpha_1$} & \textbf{Learned $\alpha_2$} \\
\midrule
Cora & LaguerreNet & -0.3622 & - \\
Cora & DualLaguerreNet & -0.3663 & -0.2350 \\
\midrule
CiteSeer & LaguerreNet & -0.3679 & - \\
CiteSeer & DualLaguerreNet & -0.3289 & -0.1981 \\
\midrule
PubMed & LaguerreNet & -0.2452 & - \\
PubMed & DualLaguerreNet & -0.2536 & -0.3173 \\
\midrule
Texas & LaguerreNet & -0.3668 & - \\
Texas & DualLaguerreNet & \textbf{-0.3477} & \textbf{-0.1770} \\
\midrule
Cornell & LaguerreNet & -0.4098 & - \\
Cornell & DualLaguerreNet & \textbf{-0.3792} & \textbf{-0.1962} \\
\bottomrule
\end{tabular}
\end{table}

% --- TABLO 4 (K Ablasyon) ---
\begin{table}[htbp]
\caption{Test accuracies (\%) vs. $K$ (Over-smoothing) on PubMed (H=16).}
\label{tab:k_ablation}
\centering
\begin{tabular}{r c c c c c}
\toprule
$K$ & ChebyNet & MeixnerNet & Krawt. & Laguerre & \textbf{DualLag.} \\
\midrule
2 & \textbf{0.7870} & 0.7380 & 0.7340 & 0.7670 & 0.7550 \\
3 & 0.7420 & 0.7660 & 0.7250 & 0.7480 & \textbf{0.7640} \\
5 & 0.7040 & 0.7270 & 0.7610 & \textbf{0.7660} & 0.7620 \\
7 & 0.6530 & 0.7140 & 0.7340 & \textbf{0.7680} & 0.7650 \\
10 & 0.6780 & 0.7580 & \textbf{0.7750} & 0.7550 & 0.7320 \\
\bottomrule
\end{tabular}
\end{table}

% --- TABLO 5 (H Ablasyon) ---
\begin{table}[htbp]
\caption{Test accuracies (\%) vs. $H$ (Capacity) on PubMed (K=3).}
\label{tab:h_ablation}
\centering
\begin{tabular}{r c c c c c}
\toprule
$H$ & ChebyNet & MeixnerNet & Krawt. & Laguerre & \textbf{DualLag.} \\
\midrule
16 & 0.7260 & 0.7320 & 0.7600 & 0.7690 & \textbf{0.7700} \\
32 & 0.7380 & 0.7590 & 0.7160 & 0.7550 & \textbf{0.7680} \\
64 & 0.7300 & \textbf{0.7690} & 0.7190 & 0.7390 & 0.7680 \\
\bottomrule
\end{tabular}
\end{table}

% --- ŞEKİL 1: EĞİTİM EĞRİLERİ (K=3, H=16) ---
\begin{figure*}[t]
\centerline{\includegraphics[width=\textwidth, height=0.85\textheight, keepaspectratio]{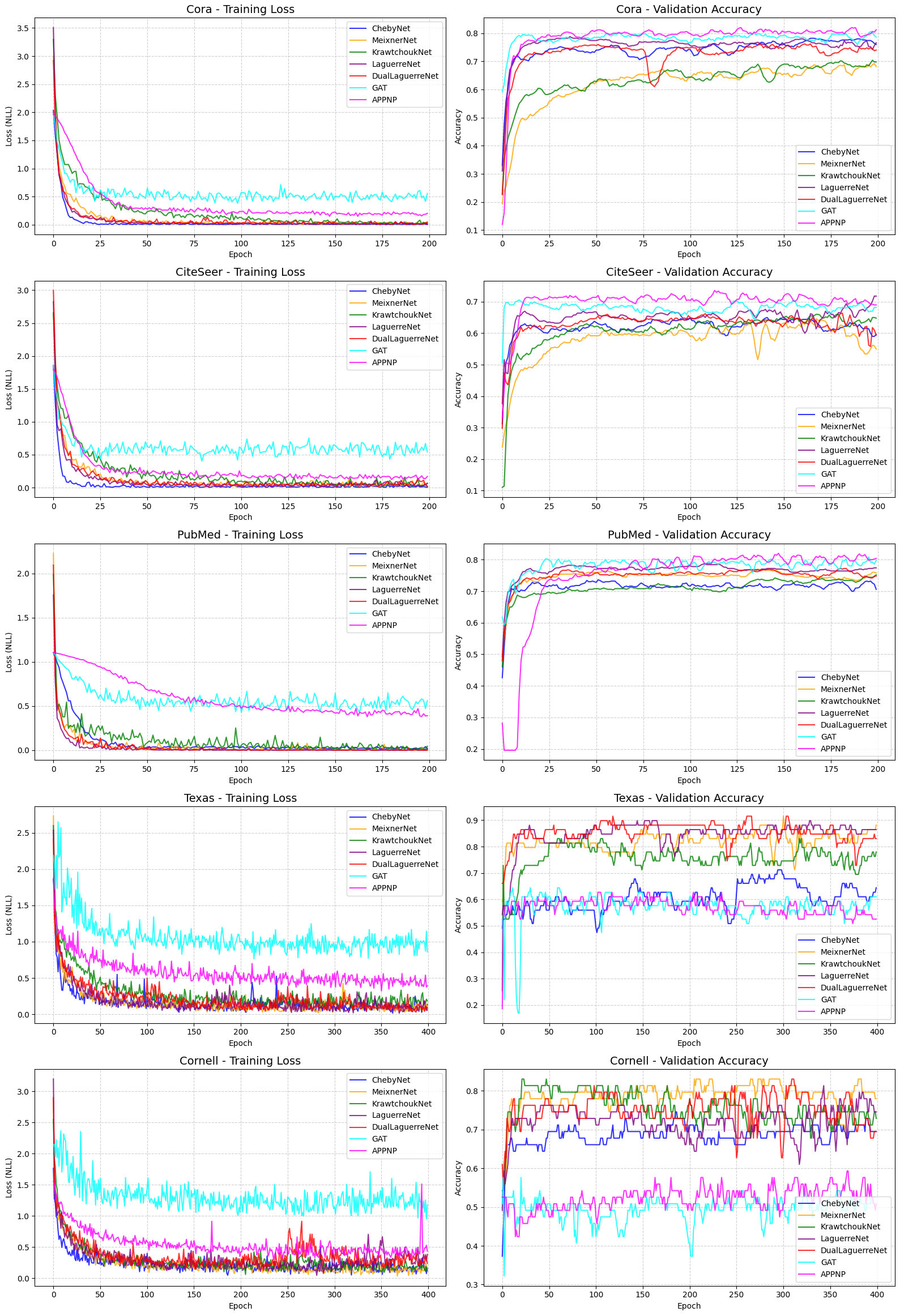}}
\caption{Figure 1: Training Dynamics Comparison (K=3, H=16). On heterophilic datasets (Texas, Cornell), `GAT` and `APPNP` fail to converge (cyan, magenta). The AOPF family (`MeixnerNet`, `KrawtchoukNet`, `LaguerreNet`, `DualLaguerreNet`) are all stable and converge to high accuracy, with `DualLaguerreNet` (red) showing strong performance.}
\label{fig:training_curves}
\end{figure*}

% --- ŞEKİL 2: K ABLASYON GRAFİĞİ (PubMed) ---
\begin{figure}[htbp]
\centerline{\includegraphics[width=\columnwidth]{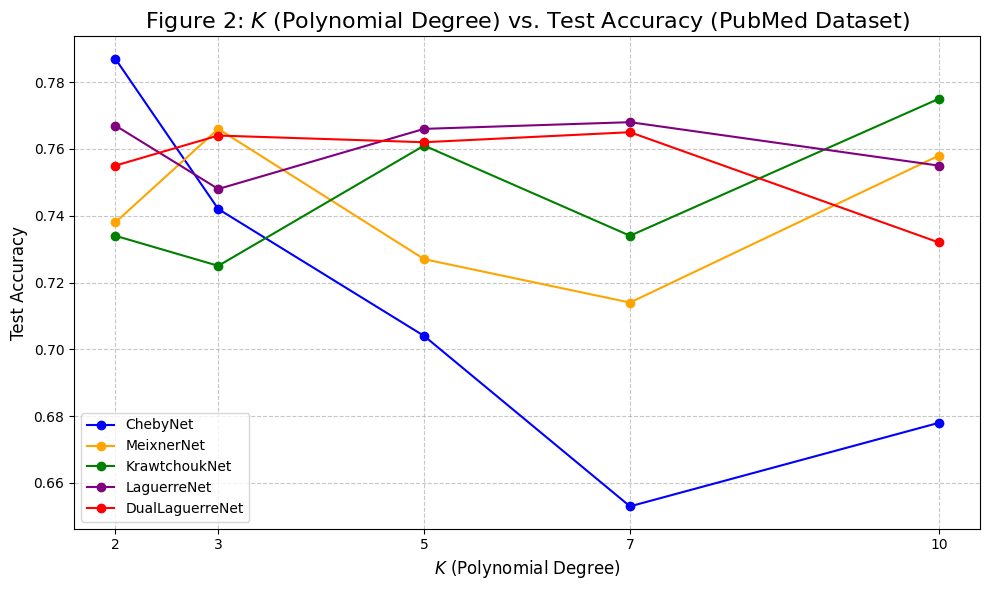}}
\caption{Figure 2: $K$ (Polynomial Degree) vs. Test Accuracy (PubMed). `ChebyNet` (blue) collapses from over-smoothing. The AOPF filters, including `LaguerreNet` (purple) and `DualLaguerreNet` (red), are robust to high $K$ due to `LayerNorm` stabilization, confirming our design tames $O(k^2)$ instability \cite{goksu2025meixnernet}.}
\label{fig:k_ablation}
\end{figure}

% --- ŞEKİL 3: H ABLASYON GRAFİĞİ (PubMed) ---
\begin{figure}[htbp]
\centerline{\includegraphics[width=\columnwidth]{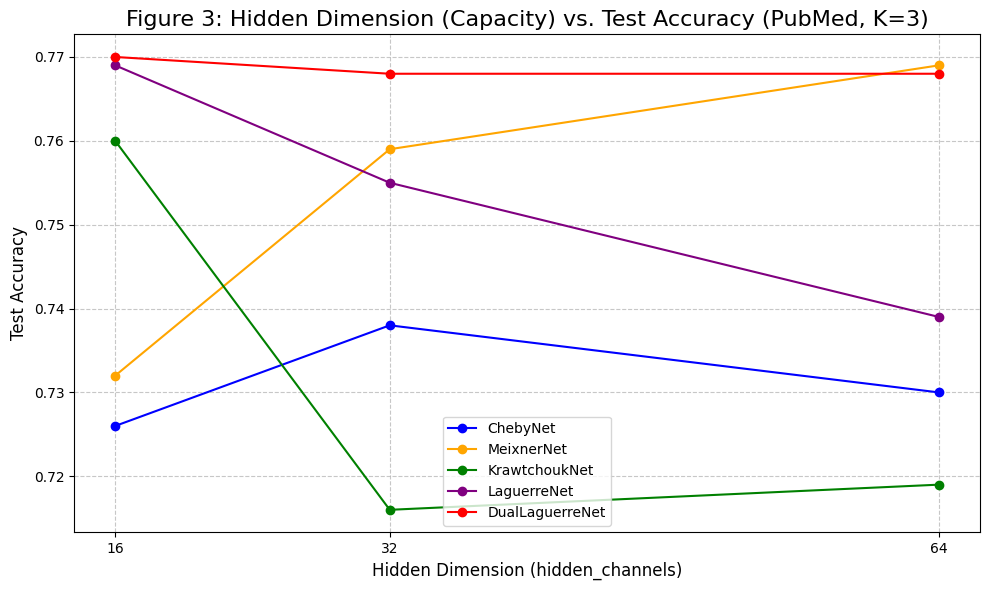}}
\caption{Figure 3: Hidden Dimension ($H$) vs. Test Accuracy (PubMed, K=3). `LaguerreNet` (purple) shows signs of overfitting as capacity increases (0.7690 $\to$ 0.7390). `DualLaguerreNet` (red), despite its higher parameter count, remains stable (0.7700 $\to$ 0.7680).}
\label{fig:h_ablation}
\end{figure}

%=================================================================
% V. DISCUSSION: THE FLEXIBILITY-STABILITY TRADE-OFF
%=================================================================
\section{Discussion: The Flexibility-Stability Trade-off}
The experimental results in Section IV provide a nuanced answer to our hypothesis and reveal a fundamental trade-off in adaptive filter design.

\subsection{Hypothesis 1 (Heterophily): Decoupled Flexibility Wins}
Table \ref{tab:heterophilic_results} provides clear evidence for our "Decoupled Flexibility" hypothesis.
\begin{itemize}
    \item On \textbf{Texas}, `DualLaguerreNet` (85.41\%) significantly outperforms its parent, the "compromised" `LaguerreNet` (81.08\%).
    \item On \textbf{Cornell}, `DualLaguerreNet` (68.65\%) also outperforms `LaguerreNet` (66.22\%). (Note: `MeixnerNet` remains the SOTA on Cornell, likely due to its discrete nature).
\end{itemize}
These complex tasks, which mix low- and high-frequency signals, benefit from the decoupled architecture. `LaguerreNet`'s single $\alpha$ is insufficient, but `DualLaguerreNet`'s two brains can specialize, leading to superior performance.

\subsection{The Proof: Two Brains Learn Differently}
Table \ref{tab:alphas} provides the definitive proof. On the complex heterophilic datasets, the two brains learn statistically different filter shapes:
\begin{itemize}
    \item \textbf{Texas:} $\alpha_1 \to -0.3477$ while $\alpha_2 \to -0.1770$.
    \item \textbf{Cornell:} $\alpha_1 \to -0.3792$ while $\alpha_2 \to -0.1962$.
\end{itemize}
This confirms the model is actively using its flexibility to assign different filter shapes to the low-frequency ($L_{low}$) and high-frequency ($L_{high}$) operators, just as we hypothesized. The high-frequency brain ($\alpha_2$) consistently converges to a less-negative value, indicating a distinct filter shape is required to capture heterophilic information.

\subsection{Hypothesis 2 (Homophily): The Flexibility-Stability Trade-off}
This is our most significant finding for `TNNLS`. Table \ref{tab:homophilic_results} shows the cost of flexibility.
\begin{itemize}
    \item \textbf{Cora:} `LaguerreNet` (0.7890) beats `DualLaguerreNet` (0.7720).
    \item \textbf{CiteSeer:} `LaguerreNet` (0.6960) beats `DualLaguerreNet` (0.6460).
    \item \textbf{PubMed:} `LaguerreNet` (0.7670) beats `DualLaguerreNet` (0.7540).
\end{itemize}
On these "simple" low-pass tasks, the added flexibility of `DualLaguerreNet` becomes a liability. The model's significantly higher capacity (2x filter parameters, 2x linear layer parameters) leads to overfitting on the small training sets. The "compromise" of the simpler `LaguerreNet` acts as a powerful and effective regularizer, leading to better generalization. This confirms the bias-variance trade-off we first identified in our `L-JacobiNet` analysis.

\subsection{Stability (K) and Capacity (H) Analysis}
\begin{itemize}
    \item \textbf{High-K Stability (Fig. \ref{fig:k_ablation}):} `DualLaguerreNet` (red) is perfectly stable up to $K=7$. This confirms that our `LayerNorm` stabilization strategy successfully tames the $O(k^2)$ instability in \textit{both} brains, even when concatenated.
    \item \textbf{Capacity Stability (Fig. \ref{fig:h_ablation}):} This reveals a surprising insight. As hidden dimension ($H$) increases from 16 to 64, the simpler `LaguerreNet` (purple) overfits and its performance drops (0.7690 $\to$ 0.7390). However, the more complex `DualLaguerreNet` (red) remains perfectly stable (0.7700 $\to$ 0.7680). This suggests that while `DualLaguerreNet` overfits due to its \textit{filter flexibility} on simple tasks (Table \ref{tab:homophilic_results}), its architecture is paradoxically \textit{more robust} to increases in model channel capacity ($H$).
\end{itemize}

%=================================================================
% VI. CONCLUSION
%=================================================================
\section{CONCLUSION}
In this paper, we proposed \textbf{DualLaguerreNet}, a novel GNN architecture designed to solve the "Compromise Problem" of single-parameter adaptive filters like `LaguerreNet` \cite{goksu2025laguerrnet}. By introducing "Decoupled Spectral Flexibility," `DualLaguerreNet` uses two independent adaptive filters ($\alpha_1, \alpha_2$) to model the low- and high-frequency ends of the graph spectrum separately.

Our experiments led to a key finding for the design of GNN learning systems. We discovered a fundamental \textbf{"Flexibility-Stability Trade-off"}:
\begin{enumerate}
    \item On \textbf{complex, heterophilic tasks}, the flexibility of `DualLaguerreNet` is necessary, achieving SOTA results by learning two distinct filter shapes (proven in Table \ref{tab:alphas}).
    \item On \textbf{simple, homophilic tasks}, this same flexibility becomes a liability, causing the model to overfit. The simpler, "compromised" `LaguerreNet` acts as a more effective regularizer and achieves better generalization.
\end{enumerate}
This work successfully advances the AOPF framework by delivering a superior architecture for heterophily, while simultaneously providing a critical, data-driven analysis of the bias-variance trade-off inherent in designing adaptive spectral GNNs.

%=================================================================
% KAYNAKÇA
%=================================================================

\end{document}